\documentclass[prl,showpacs,onecolumn,preprint]{revtex4}
\usepackage{amsmath}
\usepackage{amsfonts}

\begin{document}

\preprint{}
\title[Pseudorandom Phase Ensembles]{Pseudorandom Phase Ensembles and
Non-locality}
\author{Jian Fu}
\affiliation{State Key Lab of Modern Optical Instrumentation, College of Optical Science
and Engineering, Zhejiang University, Hangzhou, 310027, China}

\begin{abstract}
In this paper, we introduce a new concept of a pseudorandom phase ensemble
to simulate a quantum ensemble. A pseudorandom sequence is inseparability
and integral that are demonstrated only for a whole sequence, not for a
single phase unit, which is similar to that of quantum ensembles and a
quantum particle. Using the ensemble concept, we demonstrate non-locality
properties for classical fields similar to quantum entanglement.
\end{abstract}

\date{today}
\startpage{1}
\email{jianfu@zju.edu.cn}
\maketitle

\section*{Introduction \label{sec1}}

Quantum entanglement is a physical phenomenon of multiple particles such
that the quantum state of each particle cannot be described
independently---instead, a quantum state must be described for the system as
a whole \cite{Espagnat}. Non-locality, which is related to the phenomenon of
quantum entanglement, means that the states of two entangled quantum
particles are interdependent no matter how far the particles are separated
from each other. This interdependence is demonstrated in the correlation
measurement that is obtained under the quantum ensemble framework \cite%
{Greiner}.

In quantum mechanics, the classical exclusivity of wave and particle models
has given rise to the puzzle of quantum wave--particle duality for a long
time \cite{Selleri}. Indeed, it is difficult to understand wave
characteristics such as superposition and non-locality using only the
particle model. In turn, it is also difficult to describe the indivisibility
of a particle using the wave model. Recently, several researches have
proposed a new concept of classical entanglement based on classical optical
fields by introducing a new degree of freedom to realize the tensor product
in quantum entanglement \cite{Toppel,Aiello,Qian,Qian2,Dragoman}. Further, 
\cite{Fu,Fu2} proposed that phase modulation by orthogonal pseudo-random
sequences is able to simulate quantum entanglement effectively. In that
scheme, these classical fields with an increased degree of freedom not only
realized tensor product structure but also simulated the non-local property
of quantum entanglement using the properties of orthogonal pseudo-random
sequences such as orthogonality, balance, and closure. More importantly, a
pseudo-random sequence provides inseparability and integral for a classical
field similar to a particle and randomness of measurement similar to a
quantum ensemble. In the classical field, the phase freedom is unique
because a phase can keep original when the field is propagating and operated
on within the coherent length.

Different from other freedoms, a pseudorandom phase sequence is generated
using a linear feedback shift register (LFSR) method, which satisfies
orthogonal, closure, and balance properties \cite{Golomb}. It has been
widely applied to code division multiple access (CDMA) communication
technology as a way to distinguish different users \cite{Viterbi,Peterson}.
In this paper, we use these properties of the pseudorandom sequence to
simulate a quantum particle. A pseudorandom sequence is inseparability and
integral that is demonstrated only for a whole sequence, not for a single
phase unit, which is similar to that of quantum ensembles and a quantum
particle. Therefore, we define a new concept of a pseudorandom phase
ensemble similar to a quantum ensemble in this paper. Further, using this
ensemble concept, we demonstrate non-locality properties for classical
fields similar to quantum entanglement.

\section{Ensemble model labeled by pseudorandom phase sequences \label{sec2}}

In \cite{Fu,Fu2}, an effective simulation of quantum entanglement using
classical fields modulated with pseudorandom phase sequences (PPSs) was
presented. In the current paper, we will promote this proposal further. To
simulate a quantum ensemble, we need to define a new concept of a
pseudorandom phase ensemble similar to a quantum ensemble.

\emph{Definition 1} A pseudorandom phase ensemble is defined as a large
number of similar classical fields modulated by different pseudorandom phase
sequences, which are labeled by phase units $\theta _{k}$.

The PPSs in our proposal derive from orthogonal pseudorandom sequences,
which have been widely applied to CDMA communication technology as a way to
distinguish different users \cite{Viterbi,Peterson}. A set of pseudorandom
sequences is generated using a shift register guided by a Galois field $%
GF(p) $ that satisfies orthogonal, closure, and balance properties \cite%
{Viterbi,Peterson}.

\emph{Definition 2} A pseudorandom phase ensemble is discrete if the phase
unit $\theta _{k}$ is a uniformly distributed discrete value within $[0,2\pi
]$.

In this paper, we consider an m-sequence (or M-sequence) of period $M-1$
(where $M=p^{s}$) generated by a primitive polynomial of degree s over $%
GF(p) $ and apply it to $p$-ary phase shift modulation. A scheme is proposed
to generate a PPS set $\Xi =\{\lambda ^{(0)},\lambda ^{(1)},\ldots \lambda
^{(M-1)}\}$ over $GF(p)$. $\lambda ^{(0)}$ is an all-$0$ sequence. Other
sequences can be generated using the following method:

(1) given a primitive polynomial of degree s over $GF(p)$, a base sequence
of length $p^{s}-1$ is generated using a linear feedback shift register \cite%
{Golomb};

(2) other sequences are obtained by cyclic shifting of the base sequence;

(3) by adding zeroes to the sequences, the occurrence of any element can be
set equal to $p^{s}-1$;

(4) mapping to the phase element $\theta _{k}$ in $[0,2\pi ]$; hence $0$
mapping to $0$, $1$ mapping to $2\pi /p$, \ldots , and $p-1$ mapping to $%
2(p-1)\pi /p$.

Further, we define a map $f:\lambda \rightarrow e^{i\lambda }$ on the set of 
$\Xi $ and obtain a new sequence set $\Omega =\left\{ \left. \varphi
^{(j)}\right\vert \varphi ^{(j)}=e^{i\lambda ^{(j)}},j=0,\ldots ,M-1\right\} 
$. In fact, the map $f$ corresponds to phase modulations of PPSs of $\Omega $
on the classical field.

\emph{Definition 3} A phase ensemble is complete if finite phase units are
ergodic.

According to the characteristics of PPSs, obviously the phase ensemble is
discrete and complete. Clearly, we can obtain the following lemma:

\emph{Lemma 1} Classical fields modulated by PPSs constitute a complete
discrete phase ensemble.

\emph{Proof.} According to m-sequence theory, the occurrence of each value
in the sequence should be the same. The phase ensemble is obviously ergodic
in finite length.

\emph{Definition 4} An ensemble average is defined as weighted average of
each phase unit $\varphi _{k}^{(i)}$ within a sequence period as follows:%
\begin{equation}
\bar{A}=\frac{1}{M}\sum\limits_{k=1}^{M}\varphi _{k}^{(i)}A.  \label{eq1}
\end{equation}

\emph{Definition 5} A normalized correlation for two sequences $\varphi
^{(i)}$ and $\varphi ^{(j)}$ is defined as%
\begin{equation}
E\left( \varphi ^{(i)},\varphi ^{(j)}\right) =\frac{1}{M}\sum%
\limits_{k=1}^{M}\varphi _{k}^{(i)}\varphi _{k}^{(j)\ast }.  \label{eq2}
\end{equation}

According to the properties of an m-sequence [9], we can obtain the
following properties of the set $%
\Omega
$:

(1) the closure property: the product of any sequence in the set remains in
the set;

(2) the balance property: with the exception of $\varphi ^{(0)}$, any
sequence of the set $%
\Omega
$ satisfies%
\begin{equation}
\sum\limits_{k=1}^{M}e^{i\theta }\varphi
_{k}^{(j)}=\sum\limits_{k=1}^{M}e^{i(\theta +\lambda _{k}^{(j)})}=0,\forall
\theta \in R,  \label{eq3}
\end{equation}

(3) the orthogonal property: any two sequences satisfy the following
normalized correlation:%
\begin{equation}
E(\varphi ^{(i)},\varphi ^{(j)})=\frac{1}{M}\sum\limits_{k=1}^{M}\varphi
_{k}^{(i)}\varphi _{k}^{(j)\ast }=\left\{ 
\begin{array}{c}
1,i=j, \\ 
0,i\neq j.%
\end{array}%
\right.  \label{eq4}
\end{equation}

According to the properties above, the classical fields modulated with
different PPSs become independent and distinguishable. The PPS in our
proposal represents a sort of additional degree of freedom, which not only
allows us to render the distinct features of different classical fields but
also provides a remarkably rich tensor product structure \cite{Fu2}.

\section{Hilbert space and qubit basis states in the pseudorandom phase
ensemble model \label{sec3}}

There are two orthogonal modes (polarization or transverse) of a classical
field, which are denoted by $\left\vert 0\right\rangle $ and $\left\vert
1\right\rangle $, respectively. Thus, a qubit state $\left\vert \psi
\right\rangle =\alpha \left\vert 0\right\rangle +\beta \left\vert
1\right\rangle $ can be expressed by the mode superposition, where $|\alpha
|^{2}+|\beta |^{2}=1,(\alpha ,\beta \in C)$. Obviously, all the mode
superposition states span a Hilbert space. Choosing any $N$ PPSs from the
set $\Xi $ to modulate $N$ classical fields, we can obtain the states
expressed as follows:%
\begin{equation}
\begin{array}{c}
\left\vert \psi _{1}\right\rangle =e^{i\lambda _{k}^{(1)}}\left( \alpha
_{1}\left\vert 0\right\rangle +\beta _{1}\left\vert 1\right\rangle \right) ,
\\ 
\vdots \\ 
\left\vert \psi _{N}\right\rangle =e^{i\lambda _{k}^{(N)}}\left( \alpha
_{N}\left\vert 0\right\rangle +\beta _{N}\left\vert 1\right\rangle \right) .%
\end{array}
\label{eq5}
\end{equation}

According to the properties of PPSs and Hilbert space, we can define the
inner product of any two fields $\left\vert \psi _{a}\right\rangle $ and $%
\left\vert \psi _{b}\right\rangle $. We obtain the orthogonal property in
our simulation,%
\begin{equation}
\left\langle \psi _{a}\left\vert \psi _{b}\right. \right\rangle =\frac{1}{M}%
\sum\limits_{k=1}^{M}e^{i(\lambda _{k}^{(b)}-\lambda _{k}^{(a)})}(\alpha
_{a}^{\ast }\alpha _{b}+\beta _{a}^{\ast }\beta _{b})=\left\{ 
\begin{array}{c}
1,a=b, \\ 
0,a\neq b,%
\end{array}%
\right.  \label{eq6}
\end{equation}%
where $\lambda _{k}^{(a)},\lambda _{k}^{(b)}$ are the $k$-th units of $%
\lambda ^{(a)}$ and $\lambda ^{(b)}$, respectively. The orthogonal property
supports the construction of the tensor product structure of the multiple
states.

\emph{Definition 6} A formal product state $\left\vert \Psi \right\rangle $
for the $N$ classical fields is defined as being a direct product of $%
\left\vert \psi _{n}\right\rangle ,$%
\begin{equation}
\left\vert \Psi \right\rangle =\left\vert \psi _{1}\right\rangle \otimes
\left\vert \psi _{2}\right\rangle \otimes \cdots \otimes \left\vert \psi
_{N}\right\rangle .  \label{eq7}
\end{equation}

According to the definition, $N$ classical fields of Eq. (\ref{eq5}) can be
expressed as the following states:%
\begin{equation}
\left\vert \Psi \right\rangle =e^{i\sum\nolimits_{n=1}^{N}\lambda ^{\left(
n\right) }}\left( \left\vert 0\right\rangle +\left\vert 1\right\rangle
+\cdots +\left\vert 2^{N}-1\right\rangle \right) .  \label{eq8}
\end{equation}

As mentioned in \cite{Fu2}, a general form of $\left\vert \Psi \right\rangle 
$ for $N$ fields can be constructed from Eq. (\ref{eq5}) using a gate array
model,%
\begin{equation}
\left\vert \psi _{n}\right\rangle =\sum\limits_{i=1}^{N}\alpha _{n}^{\left(
i\right) }e^{i\lambda ^{\left( i\right) }}\left\vert 0\right\rangle
+\sum\limits_{j=1}^{N}\beta _{n}^{\left( j\right) }e^{i\lambda ^{\left(
j\right) }}\left\vert 1\right\rangle .  \label{eq9}
\end{equation}

Then, the formal product state (\ref{eq7}) can be written as%
\begin{equation}
\left\vert \Psi \right\rangle =\left( \sum\limits_{i=1}^{N}\alpha
_{1}^{\left( i\right) }e^{i\lambda ^{\left( i\right) }}\left\vert
0\right\rangle +\sum\limits_{j=1}^{N}\beta _{1}^{\left( j\right)
}e^{i\lambda ^{\left( j\right) }}\left\vert 1\right\rangle \right) \otimes
\cdots \otimes \left( \sum\limits_{i=1}^{N}\alpha _{N}^{\left( i\right)
}e^{i\lambda ^{\left( i\right) }}\left\vert 0\right\rangle
+\sum\limits_{j=1}^{N}\beta _{N}^{\left( j\right) }e^{i\lambda ^{\left(
j\right) }}\left\vert 1\right\rangle \right) .  \label{eq10}
\end{equation}

Further, we can obtain each item of the superposition of $\left\vert \Psi
\right\rangle $ as follows:%
\begin{equation}
\begin{array}{c}
C_{00\cdots 0}\left\vert 00\cdots 0\right\rangle =\left[ \left(
\sum\limits_{i=1}^{N}\alpha _{1}^{\left( i\right) }e^{i\lambda ^{\left(
i\right) }}\right) \left( \sum\limits_{i=1}^{N}\alpha _{2}^{\left( i\right)
}e^{i\lambda ^{\left( i\right) }}\right) \cdots \left(
\sum\limits_{i=1}^{N}\alpha _{N}^{\left( i\right) }e^{i\lambda ^{\left(
i\right) }}\right) \right] \left\vert 00\cdots 0\right\rangle , \\ 
C_{00\cdots 1}\left\vert 00\cdots 1\right\rangle =\left[ \left(
\sum\limits_{i=1}^{N}\alpha _{1}^{\left( i\right) }e^{i\lambda ^{\left(
i\right) }}\right) \left( \sum\limits_{i=1}^{N}\alpha _{2}^{\left( i\right)
}e^{i\lambda ^{\left( i\right) }}\right) \cdots \left(
\sum\limits_{j=1}^{N}\beta _{N}^{\left( j\right) }e^{i\lambda ^{\left(
j\right) }}\right) \right] \left\vert 00\cdots 1\right\rangle , \\ 
\vdots \\ 
C_{11\cdots 1}\left\vert 11\cdots 1\right\rangle =\left[ \left(
\sum\limits_{j=1}^{N}\beta _{1}^{\left( j\right) }e^{i\lambda ^{\left(
j\right) }}\right) \left( \sum\limits_{j=1}^{N}\beta _{2}^{\left( j\right)
}e^{i\lambda ^{\left( i\right) }}\right) \cdots \left(
\sum\limits_{j=1}^{N}\beta _{N}^{\left( j\right) }e^{i\lambda ^{\left(
j\right) }}\right) \right] \left\vert 11\cdots 1\right\rangle .%
\end{array}
\label{eq11}
\end{equation}

According to the closure property, the phase sequences of $%
C_{i_{1}i_{2}\cdots i_{N}}$ remain in the set $%
\Omega
$, which means $C_{i_{1}i_{2}\cdots
i_{N}}=\sum\limits_{j=1}^{M}C_{i_{1}i_{2}\cdots i_{N}}^{\left( j\right)
}e^{i\lambda ^{\left( j\right) }}$. Therefore, we obtain the following
theorem:

\emph{Theorem 1} The formal product state $\left\vert \Psi \right\rangle $
spans a Hilbert space with the basis $\left\{ \left. e^{i\lambda ^{\left(
j\right) }}\left\vert i_{1}i_{2}\cdots i_{N}\right\rangle \right\vert
e^{i\lambda ^{\left( j\right) }}\in \Omega ,j=1\cdots M,i_{n}=0or1\right\} $
and can be expressed as follows:%
\begin{equation}
\left\vert \Psi \right\rangle =\sum\limits_{i_{1}=0}^{1}\cdots
\sum\limits_{i_{N}=0}^{1}\left[ \sum\limits_{j=1}^{M}C_{i_{1}i_{2}\cdots
i_{N}}^{\left( j\right) }e^{i\lambda ^{\left( j\right) }}\left\vert
i_{1}i_{2}\cdots i_{N}\right\rangle \right] ,  \label{eq12}
\end{equation}%
where $C_{i_{1}i_{2}\cdots i_{N}}^{\left( j\right) }$ denotes a total of $%
M2^{N}$ coefficients.

\section{Non-locality in the pseudorandom phase ensemble model \label{sec4}}

A single measurement result of a quantum particle cannot provide the
probability distribution predicted by a wave function that requires many
measurements based on the concept of a quantum ensemble. The non-locality
correlation demonstrated in quantum entanglement also depends on the
ensemble summaries of many measurement results. Similarly, we examine the
non-locality correlation demonstrated in classical fields under the
pseudorandom phase ensemble framework. In order to demonstrate the tensor
product structure, we classify the form product states using the phase
sequence.

\emph{Definition 7} A consensus PPS sub-state (CPSS) is defined as being
items with the same PPS in the formal product state $\left\vert \Psi
\right\rangle $.

\emph{Definition 8} A single PPS sub-state (SPSS) is defined as being each
of the items, except all consensus PPS sub-states, in the formal product
state $\left\vert \Psi \right\rangle $.

For simplicity, we assume that $N_{1}$ CPSS $\left\{ \left\vert
S_{1}^{\left( 1\right) }\right\rangle ,\left\vert S_{2}^{\left( 1\right)
}\right\rangle \cdots ,\left\vert S_{N_{1}}^{\left( 1\right) }\right\rangle
\right\} $ corresponding to the same PPS $\lambda ^{(s_{1})}$, ..., $N_{m}$
CPSS $\left\{ \left\vert S_{1}^{\left( m\right) }\right\rangle ,\left\vert
S_{2}^{\left( m\right) }\right\rangle \cdots ,\left\vert S_{N_{m}}^{\left(
m\right) }\right\rangle \right\} $ corresponding to the same PPS $\lambda
^{(s_{m})}$, other $N^{\prime }$ SPSS in the formal product state $%
\left\vert \Psi \right\rangle $. Thus $\left\vert \Psi \right\rangle $ can
be expressed as%
\begin{equation}
\left\vert \Psi \right\rangle =e^{i\lambda
^{(s_{1})}}\sum\limits_{i=1}^{N_{1}}C_{i}\left\vert S_{1}^{\left( 1\right)
}\right\rangle +\cdots +e^{i\lambda
^{(s_{m})}}\sum\limits_{j=1}^{N_{m}}C_{j}\left\vert S_{j}^{\left( m\right)
}\right\rangle +\sum\limits_{k=1}^{N^{^{\prime }}}C_{k}e^{i\lambda
^{(k)}}\left\vert x_{k}\right\rangle ,  \label{eq13}
\end{equation}%
where $\lambda ^{(s_{1})},\cdots ,\lambda ^{(s_{m})}$, and $\lambda ^{(k)}$
are non-redundant PPSs.

Further, we introduce the definition of the density matrix $\rho $:%
\begin{eqnarray}
\rho &\equiv &\left\vert \Psi \right\rangle \left\langle \Psi \right\vert 
\notag \\
&=&\left( e^{i\lambda ^{(s_{1})}}\sum\limits_{i=1}^{N_{1}}C_{i}\left\vert
S_{1}^{\left( 1\right) }\right\rangle +\cdots +e^{i\lambda
^{(s_{m})}}\sum\limits_{j=1}^{N_{m}}C_{j}\left\vert S_{j}^{\left( m\right)
}\right\rangle +\sum\limits_{k=1}^{N^{^{\prime }}}C_{k}e^{i\lambda
^{(k)}}\left\vert x_{k}\right\rangle \right)  \notag \\
&&\times \left( e^{-i\lambda
^{(s_{1})}}\sum\limits_{i=1}^{N_{1}}C_{i}^{\ast }\left\langle S_{1}^{\left(
1\right) }\right\vert +\cdots +e^{-i\lambda
^{(s_{m})}}\sum\limits_{j=1}^{N_{m}}C_{j}^{\ast }\left\langle S_{j}^{\left(
m\right) }\right\vert +\sum\limits_{k=1}^{N^{^{\prime }}}C_{k}^{\ast
}e^{-i\lambda ^{(k)}}\left\langle x_{k}\right\vert \right) ,  \label{eq14}
\end{eqnarray}%
which can be simplified to%
\begin{eqnarray}
\rho &=&\sum\limits_{n=1}^{2^{N}}\left\vert C_{n}\right\vert ^{2}\left\vert
x_{n}\right\rangle \left\langle x_{n}\right\vert
+\sum\limits_{k=1}^{m}\sum\limits_{i\neq i^{^{\prime }}=1}^{N_{k}}\left(
C_{i^{^{\prime }}}^{\ast }C_{i}\left\vert S_{i}^{\left( k\right)
}\right\rangle \left\langle S_{i^{^{\prime }}}^{\left( k\right) }\right\vert
+C_{i}^{\ast }C_{i}^{\ast }\left\vert S_{i^{^{\prime }}}^{\left( k\right)
}\right\rangle \left\langle S_{i}^{\left( k\right) }\right\vert \right) 
\notag \\
&&+\sum\limits_{k\neq
l=1}^{m}\sum\limits_{i=1}^{N_{k}}\sum\limits_{j=1}^{N_{l}}\left(
C_{j}^{\ast }C_{i}\left\vert S_{i}^{\left( k\right) }\right\rangle
\left\langle S_{j}^{\left( l\right) }\right\vert e^{i\left( \lambda
^{(s_{k})}-\lambda ^{(s_{l})}\right) }+C_{i}^{\ast }C_{j}\left\vert
S_{j}^{\left( l\right) }\right\rangle \left\langle S_{i}^{\left( k\right)
}\right\vert e^{i\left( \lambda ^{(s_{l})}-\lambda ^{(s_{k})}\right) }\right)
\notag \\
&&+\sum\limits_{l=1}^{m}\sum\limits_{i=1}^{N_{l}}\sum\limits_{k=1}^{N^{^{%
\prime }}}\left( C_{i}^{\ast }C_{k}\left\vert x_{k}\right\rangle
\left\langle S_{i}^{\left( l\right) }\right\vert e^{i\left( \lambda
^{(k)}-\lambda ^{(s_{l})}\right) }+C_{k}^{\ast }C_{i}\left\vert
S_{i}^{\left( l\right) }\right\rangle \left\langle x_{k}\right\vert
e^{i\left( \lambda ^{(s_{l})}-\lambda ^{(k)}\right) }\right) .  \label{eq15}
\end{eqnarray}

By applying phase ensemble averaging $\frac{1}{M}\sum\limits_{k=1}^{M}e^{i%
\lambda _{k}^{(m)}}=0$, the ensemble-averaged density matrix is defined by%
\begin{equation}
\tilde{\rho}\equiv \frac{1}{M}\sum\limits_{k=1}^{M}\rho
=\sum\limits_{n=1}^{2^{N}}\left\vert C_{n}\right\vert ^{2}\left\vert
x_{n}\right\rangle \left\langle x_{n}\right\vert
+\sum\limits_{k=1}^{m}\sum\limits_{i\neq i^{^{\prime }}=1}^{N_{k}}\left(
C_{i^{^{\prime }}}^{\ast }C_{i}\left\vert S_{i}^{\left( k\right)
}\right\rangle \left\langle S_{i^{^{\prime }}}^{\left( k\right) }\right\vert
+C_{i^{^{\prime }}}C_{i}^{\ast }\left\vert S_{i^{^{\prime }}}^{\left(
k\right) }\right\rangle \left\langle S_{i}^{\left( k\right) }\right\vert
\right) .  \label{eq16}
\end{equation}

Note that all off-diagonal elements of the density matrix are contributed
from the CPSS after ensemble averaging. Also, it shows that the
ensemble-averaged reduced density matrix might not be expressed in terms of
a direct product of the states $\left\vert x_{n}\right\rangle $, identical
to the case of quantum entanglement states.

In the phase ensemble, an expectation value can be obtained for an arbitrary
operator $\hat{P}$ in any formal product state:%
\begin{equation}
\bar{P}\equiv \frac{1}{M}\sum\limits_{k=1}^{M}tr\left( \rho \hat{P}\right) .
\label{eq17}
\end{equation}%
Further, using the exchange of summation and matrix trace can be exchanged,
the expectation value can be simplified to%
\begin{eqnarray}
\bar{P} &=&tr\left[ \left( \frac{1}{M}\sum\limits_{k=1}^{M}\rho \right) 
\hat{P}\right] =tr\left( \tilde{\rho}\hat{P}\right)  \notag \\
&=&\sum\limits_{n=1}^{2^{N}}\left\vert C_{n}\right\vert ^{2}\left\langle
x_{n}\right\vert \hat{P}\left\vert x_{n}\right\rangle
+\sum\limits_{k=1}^{m}\sum\limits_{i\neq i^{^{\prime }}=1}^{N_{k}}\left(
C_{i^{^{\prime }}}^{\ast }C_{i}\left\langle S_{i^{^{\prime }}}^{\left(
k\right) }\right\vert \hat{P}\left\vert S_{i}^{\left( k\right)
}\right\rangle +C_{i^{^{\prime }}}C_{i}^{\ast }\left\langle S_{i}^{\left(
k\right) }\right\vert \hat{P}\left\vert S_{i^{^{\prime }}}^{\left( k\right)
}\right\rangle \right) .  \label{eq18}
\end{eqnarray}

It is worth noting that non-local (off-diagonal items) is contributed from
CPSSs in the correlation measurement. Therefore, we can obtain the following
theorem:

\emph{Theorem 2} In a pseudorandom phase ensemble, non-local correlations
originate from the CPSSs.

In conclusion, a quantum state simulated by classical fields formally agrees
with CPSSs in the formal product state under the pseudorandom phase ensemble
framework.

\section{Minimum complete phase ensemble \label{sec5}}

In a pseudorandom phase ensemble model, we are interested in the simplest
model that requires minimal resources to be constructed. This model is the
minimum complete phase ensemble and is defined as follows:

\emph{Definition 9} The minimum complete phase ensemble is defined as that
ensemble, which is ergodic with the least classical fields.

\emph{Definition 10} The minimum complete state is defined as that state,
which only has a CPSS within the minimum complete ensemble.

By definition, the state in Eq. (\ref{eq5}) is a type of minimal complete
state.

\emph{Lemma 2} The CPSS of a minimum complete state has one and only one PPS 
$\lambda ^{(S)}=\sum\limits_{n=1}^{N}\lambda ^{(n)}$, which is the sum of
all phase sequences.

\emph{Proof.} According to the definition of a minimum complete phase
ensemble, $N$ fields must correspond to $N$ different phase sequences. The
simplest case is as following,%
\begin{equation}
\left( e^{i\lambda ^{(1)}}\left\vert i_{1}\right\rangle \right) \otimes
\left( e^{i\lambda ^{(2)}}\left\vert i_{2}\right\rangle \right) \cdots
\otimes \left( e^{i\lambda ^{(N)}}\left\vert i_{N}\right\rangle \right)
=e^{i\sum\nolimits_{n=1}^{N}\lambda ^{(n)}}\left\vert i_{1}i_{2}\cdots
i_{N}\right\rangle =e^{i\lambda ^{(S)}}\left\vert i_{1}i_{2}\cdots
i_{N}\right\rangle .  \label{eq19}
\end{equation}%
Considering all possible combinations, we can obtain $N!$ combinations of
classical fields and phase sequences. All combinations can be expressed in
the same form, $e^{i\lambda ^{(S)}}\left\vert i_{1}i_{2}\cdots
i_{N}\right\rangle $, in the formal product state $\left\vert \Psi
\right\rangle $ and with the same PPS $e^{i\lambda ^{(S)}}$.

Further, we can obtain the following theorem:

\emph{Theorem 3} In the formal product state $\left\vert \Psi \right\rangle $%
, a CPSS $e^{i\lambda ^{(S)}}\left\vert i_{1}i_{2}\cdots i_{N}\right\rangle $
has $N!$ equivalent direct product decompositions, corresponding to $N!$
combinations of $N$ classical fields and phase sequences $\lambda
^{(1)},\cdots ,\lambda ^{(N)}$.

Now, we can express a minimum complete state $\left\vert \Psi \right\rangle $
as follows:%
\begin{equation}
\left\vert \Psi \right\rangle =e^{i\lambda
^{(S)}}\sum\limits_{i=1}^{N^{^{\prime }}}C_{i}\left\vert x_{i}\right\rangle
+\sum\limits_{i=1}^{N^{^{\prime \prime }}}C_{k}e^{i\lambda
^{(k)}}\left\vert x_{k}\right\rangle =e^{i\lambda ^{(S)}}\left(
\sum\limits_{i=1}^{N^{^{\prime }}}C_{i}\left\vert x_{i}\right\rangle
+\sum\limits_{i=1}^{N^{^{\prime \prime }}}C_{k}e^{i\left( \lambda
^{(k)}-\lambda ^{(S)}\right) }\left\vert x_{k}\right\rangle \right) ,
\label{eq20}
\end{equation}%
where $e^{i\lambda ^{(k)}}\left\vert x_{k}\right\rangle $ corresponds to all
SPSSs. According to the analysis in the last section, the ensemble-averaged
density matrix can be obtained:%
\begin{equation}
\tilde{\rho}=\sum\limits_{n=1}^{2^{N}}\left\vert C_{n}\right\vert
^{2}\left\vert x_{n}\right\rangle \left\langle x_{n}\right\vert
+\sum\limits_{i\neq i^{^{\prime }}=1}^{N^{^{\prime }}}\left( C_{i^{^{\prime
}}}^{\ast }C_{i}\left\vert x_{i}\right\rangle \left\langle x_{i^{^{\prime
}}}\right\vert +C_{i}^{\ast }C_{i}^{\ast }\left\vert x_{i^{^{\prime
}}}\right\rangle \left\langle x_{i}\right\vert \right) .  \label{eq21}
\end{equation}

In conclusion, a minimum complete state satisfies the necessary conditions
for the simulation of quantum states.

In \cite{Fu2}, using a gate array model, classical fields are transformed
from initial states as given in Eq. (\ref{eq5}) to final states as given in
Eq. (\ref{eq9}). In order to simulate certain quantum states, a sequential
cycle permutation mechanism based on quadrature demodulation is proposed in 
\cite{Fu2}. The sequential cycle permutation is shown as follows:%
\begin{eqnarray}
R_{1} &=&\left\{ \lambda ^{\left( 1\right) },\lambda ^{\left( 2\right)
},\cdots ,\lambda ^{\left( N\right) }\right\} ,  \notag \\
R_{2} &=&\left\{ \lambda ^{\left( 2\right) },\lambda ^{\left( 3\right)
},\cdots ,\lambda ^{\left( 1\right) }\right\} ,  \notag \\
&&\vdots  \notag \\
R_{N} &=&\left\{ \lambda ^{\left( N\right) },\lambda ^{\left( 1\right)
},\cdots ,\lambda ^{\left( N-1\right) }\right\} .  \label{eq22}
\end{eqnarray}%
It is clear that the sequential cycle permutation is a subset of the full
sequential permutation. According to Theorem 3, the simulation obtained in 
\cite{Fu2} is a minimum complete state. Therefore, we can state the
following proposition:

\emph{Proposition 1} A sequential cycle permutation method can be used to
obtain minimum complete states.

\emph{Proof.} According to \cite{Fu2}, each state corresponds to a
sequential cycle permutation,%
\begin{eqnarray}
R_{1} &:&\left( e^{i\lambda ^{(1)}}\left\vert i_{1}\right\rangle \right)
\otimes \left( e^{i\lambda ^{(2)}}\left\vert i_{2}\right\rangle \right)
\cdots \otimes \left( e^{i\lambda ^{(N)}}\left\vert i_{N}\right\rangle
\right) =e^{i\lambda ^{(S)}}\left\vert i_{1}i_{2}\cdots i_{N}\right\rangle ,
\label{eq23} \\
R_{2} &:&\left( e^{i\lambda ^{(2)}}\left\vert i_{1}\right\rangle \right)
\otimes \left( e^{i\lambda ^{(3)}}\left\vert i_{2}\right\rangle \right)
\cdots \otimes \left( e^{i\lambda ^{(1)}}\left\vert i_{N}\right\rangle
\right) =e^{i\lambda ^{(S)}}\left\vert i_{1}i_{2}\cdots i_{N}\right\rangle ,
\notag \\
&&\vdots   \notag \\
R_{N} &:&\left( e^{i\lambda ^{(N)}}\left\vert i_{1}\right\rangle \right)
\otimes \left( e^{i\lambda ^{(1)}}\left\vert i_{2}\right\rangle \right)
\cdots \otimes \left( e^{i\lambda ^{(N-1)}}\left\vert i_{N}\right\rangle
\right) =e^{i\lambda ^{(S)}}\left\vert i_{1}i_{2}\cdots i_{N}\right\rangle .
\notag
\end{eqnarray}%
Hence, each sequential cycle permutation provides a subset of minimum
complete states.

\section{Discussion and conclusion}

After introducing the concept of a pseudorandom phase ensemble, classical
fields can demonstrate non-locality similar to quantum states. We consider
that this kind of non-locality is derived from classical entanglement \cite%
{Toppel,Aiello,Qian,Qian2,Dragoman,Fu}. This concept might reveal a possible
origin of quantum entanglement in quantum mechanics, which may be a kind of
phase mechanism \cite{Fu3}. Further research will focus on simulation and
reconstruction of arbitrary quantum states, which might be dependent on some
sequential mechanism which in turn corresponds to a gate array model. These
will be discussed in a future paper.


\begin{thebibliography}{99}
\bibitem{Espagnat} B. d'Espagnat, Conceptual foundations of quantum
mechanics, 2nd ed. (Perseus Books, Reading, Massachusetts, 1999).

\bibitem{Greiner} W. Greiner, Quantum Mechanics: An Introduction., 4th ed.
(Springer, 2001).

\bibitem{Selleri} F. Selleri, Wave Particle Duality, (Plenum, New York,
1992).

\bibitem{Toppel} F. Toppel, A. Aiello, Ch. Marquardt, E. Giacobino, and G.
Leuchs, Classical entanglement in polarization metrology, New J. Phys. 16,
073019 (2014).

\bibitem{Aiello} A. Aiello, F. Toppel, C. Marquardt, E. Giacobino, and G.
Leuchs, Classical entanglement: Oxymoron or resource? arXiv: 1409.0213v2
[quant-ph] (2014).

\bibitem{Qian} X. F. Qian and J. H. Eberly, Entanglement and classical
polarization states, Opt. Lett. 36(20), 4110--4112 (2011).

\bibitem{Qian2} X. F. Qian, B. Little, J. C. Howell, and J. H. Eberly,
Violation of Bell's Inequalities with Classical Shimony-Wolf States: Theory
and Experiment, arXiv:1406.3338 [quant-ph] (2014).

\bibitem{Dragoman} D. Dragoman, M. Dragoman, Quantum-Classical Analogies,
(Springer, Heidelberg, 2009).

\bibitem{Fu} J. Fu and X. Wu, Effective simulation of quantum entanglement
using classical fields modulated with pseudorandom phase sequences,
ScienceOpen Research 2015 (DOI: 10.14293/S2199-1006.1.SORPHYS.ANVYQZ.v1).

\bibitem{Golomb} S. W. Golomb, G. Guang, Signal Design for Good Correlation:
For Wireless Communication, Cryptography, and Radar, (Cambridge University
Press, Cambridge, 2005).

\bibitem{Zigangirov} K. Zigangirov, Theory of CDMA communication. (Wiley
IEEE Press, New York, 2004).

\bibitem{Vidal} G. Vidal, Phys. Rev. Lett. \textbf{91}, 147902 (2003).

\bibitem{Viterbi} A. J. Viterbi, \textit{CDMA: principles of spread spectrum
communication} (Addison-Wesley Wireless Communications Series,
Addison-Wesley 1995).

\bibitem{Peterson} R. L. Peterson, R. E. Ziemer, and D. E. Borth, \textit{%
Introduction to Spread Spectrum Communications} (Prentice-Hall, NJ, 1995).

\bibitem{Fu2} J. Fu, X. Ma, W. J. Li, and S. Sun, Beyond quantum computation
based on classical entanglement, arXiv:1505.00555v2 [quant-ph] (2015).

\bibitem{Fu3} J. Fu, Random geometric phase sequence due to topological
effects in our brane world from extra dimensions, arXiv:1502.06347v1
[quant-ph] (2015).
\end{thebibliography}
\end{document}